\documentclass[aip,
 jmp,%
 amsmath,amssymb,
preprint,%
]{revtex4-1}

\usepackage{graphicx}% Include figure files
\usepackage{dcolumn}% Align table columns on decimal point
\usepackage{bm}% bold math
\begin{document}

%\preprint{AIP/123-QED}

\title{Dimension dependence of negative differential thermal resistance in graphene nanoribbons}% Force line breaks with \\
%\thanks{Footnote to title of article.}

\author{Bao-quan  Ai}
 \affiliation{Laboratory of Quantum Information Technology,
ICMP and SPTE, South China Normal University,  Guangzhou 510006, China}%Lines break automatically or can be forced with \\

\author{Wei-rong Zhong}%
%\email{wrzhong@jnu.edu.cn}
\affiliation{Department of Physics, College of Science and Engineering,
Jinan University,  Guangzhou 510632, China
%\\This line break forced with \textbackslash\textbackslash
}%

\author{Bambi Hu}
 \affiliation{Department of Physics, University of Houston, Houston, Texas 77204-5005, USA%\\This line break forced% with \\
}%

\date{\today}% It is always \today, today,

\begin{abstract}
Negative differential thermal resistance (NDTR) in approximate graphene nanoribbons (GNRs) is investigated from one dimension to three dimensions by using classical molecular dynamics method. For single-layer GNRs, NDTR can not be observed for very narrow GNRs (one dimension), NDTR appears when the width of GNRs increases (two dimensions). However, NDTR disappears gradually on further increasing the width. For multiple-layer GNRs, when the number of the layers increases, GNRs becomes from two-dimensional system to three-dimensional system, NDTR regime reduces and eventually disappears. In addition, when the length of GNRs increases, NDTR regime also reduces and vanishes in the thermodynamic limit.
These effects may be useful for designing thermal devices where NDTR plays an important role.
\end{abstract}

%Use showkeys class option if keyword
\maketitle
\section{Introduction}
\indent Heat conduction in low-dimensional systems has recently become the
subject of a large number of theoretical and experimental studies\cite{a1}. The theoretical interest in this field lies
in the rapid progress in probing and manipulating thermal properties
of nanoscale systems, which unveils the possibility of designing
thermal devices with optimized performance at the atomic scale.
As we all know, devices that control the transport of electrons, such as the electrical diode
and transistor, have been extensively studied and led to the
widespread applications in modern electronics. However, it
is far less studied for their thermal counterparts as to control
the transport of phonons (heat flux), possibly by reason that
phonons are more difficult to control than electrons. Recently, it has been revealed by theoretical studies in model
systems that, such as electrons and photons, phonons can also perform interesting function, which shed light on the
possible designs of thermal devices\cite{a2,a3,a4,a5,a6,a7,a8,a9,a10,a11}.  The nonlinear systems with
structural asymmetry were predicted to exhibit thermal rectification\cite{a2,a3,a4,a5,a6,a7,a8,a9}, which has
triggered model designs of various types of thermal devices. Remarkably, a thermal
rectifier has been experimentally realized by using gradual
mass-loaded carbon and boron nitride nanotubes \cite{a9}.
The theoretical models of thermal transistors \cite{a6}, thermal logic gates \cite{a10}, and
thermal memory\cite{a11} are also proposed. Most of
these studies are relevant to heat conduction in the nonlinear
response regime, where the counterintuitive phenomenon of NDTR may
be observed and plays an important role in the operation of those
devices.

\indent NDTR refers to the phenomenon where the resulting heat flux
decreases as the applied temperature difference (or gradient)
increases. Usually, the studies on NDTR have been on the
models with structural inhomogeneity, such as the two-segment
Frenkel-Kontorova model \cite{a6,a12}, the weakly coupled
two-segment $\phi^{4}$ model \cite{a13}, and the anharmonic graded
mass model \cite{a5}. However, structural asymmetry is not a
necessary condition for NDTR and it can also occur in absolutely symmetric structures\cite{a14}.
Recently, Hu and coworkers \cite{a15} have found from molecular dynamics simulations that NDTR is possible in both asymmetric and symmetric GNRs. GNRs may be good candidate martials for designing thermal devices such as thermal transistors, thermal logic gates, and thermal memory. Therefore, it very necessary to find in which kind of GNRs NDTR can occur.
Recent works \cite{a16} have shown that dimension crossover can strongly affect heat transport in GNRs.
Therefore, it would be interesting to find dimension dependence of NDTR in GNRs.  In this paper, we focus on finding how the dimension of GNRs affects the  appearance of NDTR  by using nonequilibrium molecular dynamics method.

\section {Model and methods}
\indent Many theoretical models have
been proposed and applied to explain and predict the thermal properties of graphene. The experiments and theories \cite{d1} show that the carrier density of non-doped graphene is relatively low, the phonon contribution overwhelms the electronic one by orders of magnitude  and the electronic contribution to thermal conductivity (Wiedemann- Franz law) is negligible.  The thermal conductivity of graphene is thus dominated by phonon transport, namely diffusive conduction at high temperature and ballistic conduction at sufficiently low temperature. Therefore, classical molecular dynamics is widely used in calculating the thermal conductivities of GNRs.

\begin{figure}[htbp]
\begin{center}\includegraphics[height=3cm]{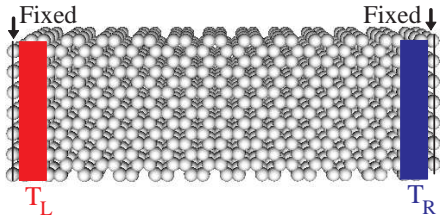}
\caption{(Color online) Schematic diagram of the armchair GNRs coupling two heat baths. The two ends along the length of GNRs are fixed and the other boundaries are free. $T_{L}=600K$ and $T_{R}$ is varied.}\label{1}
\end{center}
\end{figure}

\indent In this paper, we study thermal transport of GNRs shown in Fig. 1 by using classical molecular dynamics. For a ribbon, its length $l$ is greater than its width
$w$, which in turn is larger than the effective thickness $h$ (i.e., $l>w >h$). In the simulations, we have used the Tersoff-Brenner potential \cite{b1} for carbon-carbon
interaction in the intra-plane. For multiple-layer GNRs, van der Waals interactions between the different layers are modeled by Lennard-Jones potential \cite{b2}.

\indent The Tersoff-Brenner potential between a pair of atoms can be written as
\begin{equation}\label{}
    U_{ij}^{T}=f_{C}(r_{ij})[V_{R}(r_{ij})-b_{ij}V_{A}(r_{ij})],
\end{equation}
where $r_{ij}$ is the separation distance between two atoms $i$ and $j$. $V_{R}$ and $V_{A}$ are the repulsive and attractive Morse type potentials, respectively,
\begin{equation}\label{}
    V_{R}(r_{ij})=A\exp(-\mu r_{ij}), V_{A}(r_{ij})=B\exp(-\lambda r_{ij}),
\end{equation}
and $f_{C}$ is a smooth cutoff function with parameters $R$ and $S$,
\begin{equation}\label{}
  f_{C}(r_{ij})=\left\{
\begin{array}{ll}
   1,&\hbox{as $r_{ij}<R$};\\
   \frac{1}{2}+\frac{1}{2}\cos[\frac{\pi(r_{ij}-R)}{S-R}] ,&\hbox{as $R\leq r_{ij}\leq S$};\\
   0 ,&\hbox{as $r_{ij}>S$},\\
\end{array}
\right.
\end{equation}
$b_{ij}$ implicitly contains multiple-body information and thus the whole potential function is actually a multiple-body potential. The parameters $A$, $B$, $\mu$, $\lambda$, $S$, $R$ and the detailed information of $b_{ij}$ are given in Brenner's reference \cite{b1}.

\indent The form of the Lennard-Jones potential is shown as follows
\begin{equation}\label{}
    U_{ij}^{L}=-\frac{P}{r_{ij}^6}+\frac{Q}{r_{ij}^{12}},
\end{equation}
where $P$ and $Q$ are the attractive and repulsive constants, respectively. For graphene-graphene  interaction\cite{b2}, $P=15.2$ $eV\times {\AA}^{6}$,  $Q=24.1\times10^3$ $eV\times{\AA}^{12}$.

Therefore, the total potential between a pair of atoms is $U_{ij}=U^{T}_{ij}+U^{L}_{ij}$ and the force between two atoms can be found by taking the gradient of the
potential function with respect to their distance: $\vec{F}_{ij}=-\nabla U_{ij}$, where $\nabla$ is the gradient operator.  Then, the net force on a particular
atom can be found by summing the forces due to all other atoms in the system,
\begin{equation}\label{}
\vec{F}_{i}=\sum_{j\neq i}\vec{F}_{ij}=-\sum_{j\neq i}\nabla U_{ij}.
\end{equation}

We place atoms at the two ends of GNRs in the thermostats with temperatures $T_{L}$ (left end) and $T_{R}$
(right end) shown in Fig. 1, respectively.
The equations of motion for the atoms in the Nose-Hoover thermostats\cite{b3,b4} are
\begin{equation}\label{}
 \frac{d\vec{r}_i}{dt}=\frac{\vec{p}_i}{m},   \frac{d\vec{p}_{i}}{dt}=\vec{F}_{i}-\Gamma \vec{p}_{i}, \frac{d\Gamma}{dt}=\frac{1}{\tau^{2}}[\frac{T(t)}{T_{0}}-1],
\end{equation}
where the subscript $i$ runs over all atoms in the thermostat. $\vec{r}_{i}$ and $\vec{p}_{i}$ are the position vector and momentum of the $i$th atom, respectively.  $\vec{F}_{i}$ is the force applied on the $i$th atom which can be obtained from Eq. (5). $\Gamma$ is the dynamics parameter of the thermostat and $\tau$ is the relaxation time. $T(t)$ is the instant temperature of the thermostat at time $t$, which can be defined as $T(t)=\frac{2}{3Nk_{B}}\sum_{i}\frac{\vec{p}_{i}\cdot\vec{p}_{i}}{2m}$. $T_{0}$($=T_{L}$ or $T_{R}$) is the set temperature of the thermostat. $N$ is the number of atoms in the thermostat, $k_{B}$ is the Boltzmann constant,  and $m$ is the mass of the carbon atom. The atoms between the two thermostats are obeying the Newton's law motion,
\begin{equation}\label{}
\frac{d\vec{r}_j}{dt}=\frac{\vec{p}_j}{m},    \frac{d\vec{p}_{j}}{dt}=\vec{F}_{j},
\end{equation}
where $j$ runs over all the atoms between the two thermostats.

\indent The velocity Verlet method is employed to integrate the equations (1-7) of motion at the given initial positions and velocities of the all atoms.
The time step of $0.55fs$, and the simulation runs for $1\times 10^{8}$ time steps giving a total molecular dynamics time of $55ns$. The statistic average of interesting quantities start from half of the total time, i.e., $5\times 10^7$ time steps are used to relax the system to a stationary state. We set the relaxation time $\tau=1ps$. The distance between the neighbor layers is $0.335nm$ and the bond length of carbon-carbon is $1.48{\AA}$. In order to avoid the spurious global rotation of GNRs in the simulations, we use the fixed boundary conditions for the two ends along the length of GNRs.

\indent The heat bath acts on the atom with a force $-\Gamma \vec{p}_{i}$, thus the power of heat bath is $-\Gamma \vec{p}_{i}\cdot\vec{p}_{i}/m$, which can also be regarded as the heat flux coming out of the high temperature heat bath and injecting into the low temperature heat bath. The total heat flux (thermal current) from the heat bath to the system can be obtained  \cite{b5} by
\begin{equation}\label{}
    J=\sum_{i}[-\Gamma \vec{p}_{i}\cdot\vec{p}_{i}/m]=-3\Gamma Nk_{B}T(t),
\end{equation}
 where $i$ runs over all the atoms in the thermostat.

\indent Because thermal transport strongly depends on the phonon density of states (PDOS) which is the number of vibrational states per unit frequency, it is necessary to study the full and transversal edge PDOS of the system to understand the appearance of NDTR.  The Fourier transform of the autocorrelation function  is the PDOS \cite{pdos}.
\begin{equation}\label{}
    D(\omega)=\int_{0}^{\infty}dt\exp(2\pi\omega t)\Sigma_{j}\frac{\langle\vec{\nu}_{j}(t)\cdot\vec{\nu}_{j}(0)\rangle}{\langle\vec{\nu}_{j}(0)^{2}\rangle},
\end{equation}
where $\vec{\nu}_{j}$ represents the velocity vector of the $j$th atom, and the angle brackets denote an average over all atoms and all time windows. The transversal edge PDOS is obtained from Eq. (9) where $j$ runs over all the atoms at the transversal edge.  Similarly, full PDOS is
calculated from Eq. (9) where $j$ runs all the atoms of the GNRs.
\section{Results and discussion}

 \indent For the convenience of discussion on NDTR, thermal current can be written as $J\propto\kappa(\overline{T},\Delta T)\Delta T$, where $\kappa$ is the effective thermal conductivity which depends on $\overline{T}$ and $\Delta T$. $\Delta T =T_{L}-T_{R}$ is the temperature difference and $\overline{T}\equiv\frac{T_{L}+T_{R}}{2}=T_{L}-\frac{\Delta T}{2}$ is the average temperature. Obviously, NDTR may occur only when  the effective thermal conductivity $\kappa(\overline{T},\Delta T)$ decreases remarkably as $\Delta T$ increases.
 \begin{figure}[htbp]
\begin{center}\includegraphics[height=7cm]{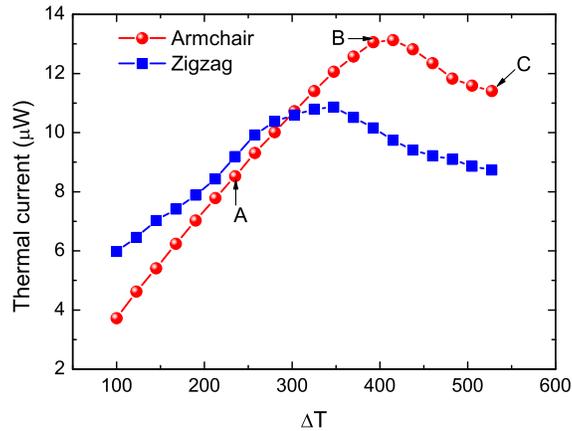}
\caption{(Color online) Thermal current $J$ as a function of the temperature difference $\Delta T$ for both armchair and zigzag GNRs. $w$ is $1.67nm$ for armchair GNRs and $1.63nm $ for zigzag GNRs. $l=5nm$, $T_{L}=600K$ and $T_{R}$ is varied. }\label{1}
\end{center}
\end{figure}

Figure 2 shows the heat transport in GNRs for both armchair and zigzag edges. When $\Delta T $ is not large, $J$ and $\Delta T$ are proportional to each other and the system is within the linear response regime. But for large values of $\Delta T$, the system enters the nonlinear response regime, where NDTR may occur. Since the results from zigzag GNRs are qualitatively similar to that from armchair GNRs,  we mainly study dimension dependence of NDTR in armchair GNRs.

 \indent Obviously,  the phenomena of NDTR  is not caused by the asymmetry, since GNRs is absolutely symmetric.  The appearance of NDTR in GNRs can be explained by the transversal edge effects of GNRs\cite{b6}.  From Eq. (9) we can obtain the transversal edge PDOS shown in Fig. 3.  It is found that there are some strong low-frequency peaks which shows the existence of the edge-localized phonon modes.
 The transversal edge of GNRs will reduce the effective thermal conductivity $\kappa$ owing to the appearance of the edge-localized phonon modes.  The edge-localized phonons can interact with other low energy phonons and thus reduce the phonons' mean free paths, which reduces the effective thermal conductivity $\kappa$.  When the average temperature increases (shown in Fig. 3), the heights of the low-frequency peaks decrease and the peaks become not significant gradually. In other words, the number of edge-localized phonon modes increases as the average temperature decreases. When the applied temperature difference $\Delta T$ increases from zero with $T_{L}=600K$, the average temperature $\overline{T}$ of the system decreases, the effect of the edge-localized phonon modes becomes more significant. At larger values of $\Delta T$ (i.e., lower values of the average temperature), the edge-localized phonon dominates the transport and the effective thermal conductivity $\kappa$ reduces remarkably, so NDTR appears.

\begin{figure}[htbp]
\begin{center}\includegraphics[height=7cm]{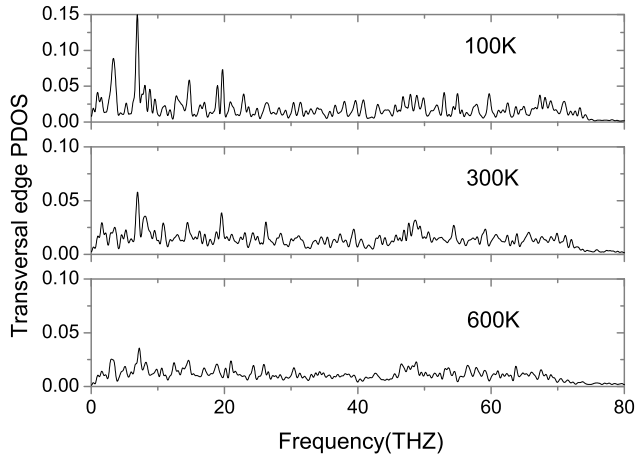}
\caption{Transversal edge PDOS (arbitrary units) of single-layer armchair GNRs for different average temperature $\overline{T}$. $w=1.67$ and $l=5nm$. }\label{1}
\end{center}
\end{figure}

\begin{figure}[htbp]
\begin{center}\includegraphics[height=7cm]{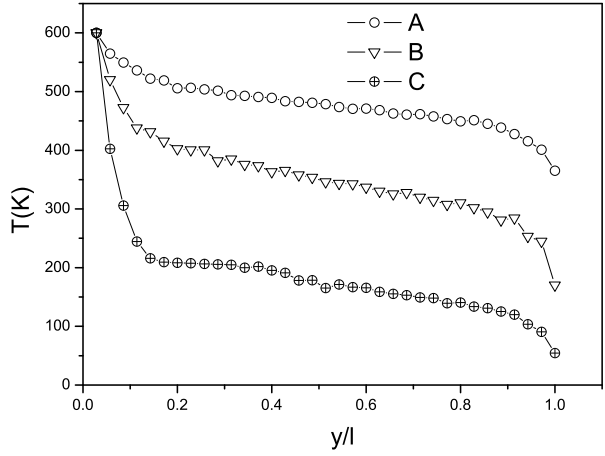}
\caption{Temperature profiles for different points $A$, $B$ and $C$ described in Fig. (2). $y/l$ is the relative position along the transport.}\label{1}
\end{center}
\end{figure}

\indent In order to understand NDTR, we also study the temperature profiles for different points $A$, $B$ and $C$ described in Fig. 2.
The results are depicted in Fig. 4.  As the applied temperature difference
$\Delta T$ increases, the system undergoes a transition from the linear
to the nonlinear response regime, with the latter being generally characterized by a nonuniform local temperature gradient.
For very large value of $\Delta T$ (e.g. point $C$), there is a big temperature jump (about $200K$) at the high temperature boundary which indicates that
the big thermal boundary resistance appears in heat transport (small thermal current). Therefore, NDTR will occur.

\begin{figure}[htbp]
\begin{center}\includegraphics[height=7cm]{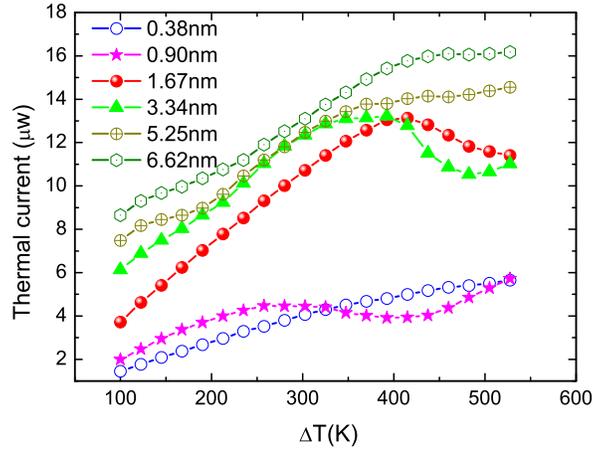}
\caption{(Color online) Thermal current $J$ as a function of the temperature difference $\Delta T$ for different widths $w$ of single-layer armchair GNRs. $w=0.38, 0.90, 1.67, 3.34, 5.25$, and $6.62nm$. $l=5nm$, $T_{L}=600K$ and $T_{R}$ is varied. }\label{1}
\end{center}
\end{figure}

\indent Figure 5 shows the relation between thermal current $J$ and temperature difference $\Delta T$ for different widths of the single-layer GNRs.  It is found that NDTR regime varies with the width $w$ of GNRs. When the width $w$ is very small, e.g., $w=0.38nm$, no NDTR can be observed. This can be understood as follows: GNRs with very small width reduces to one-dimensional atom chain\cite{b7} without on-site potentials, for example Fermi-Pasta-Ulam (FPU) chain, in this kind of chain, thermal current always increases with the applied temperature difference\cite{a14}. Therefore, NDTR can not occur for very narrow GNRs (one dimension).  When the width $w$ increases, the system changes from one dimension to two dimensions, the onset of NDTR can be observed (e.g., $w=0.90, 1.67, 3.34nm$). However, NDTR will gradually disappear on further increasing the width $w$ (e.g., $w=5.25, 6.62nm$). On further increasing $w$, the number of the total phonon modes of GNRs increases, while the number of edge-localized phonon modes does not changes. Thus the effect of edge-localized phonon modes reduces gradually and NDTR disappears.

\begin{figure}[htbp]
\begin{center}\includegraphics[height=5cm]{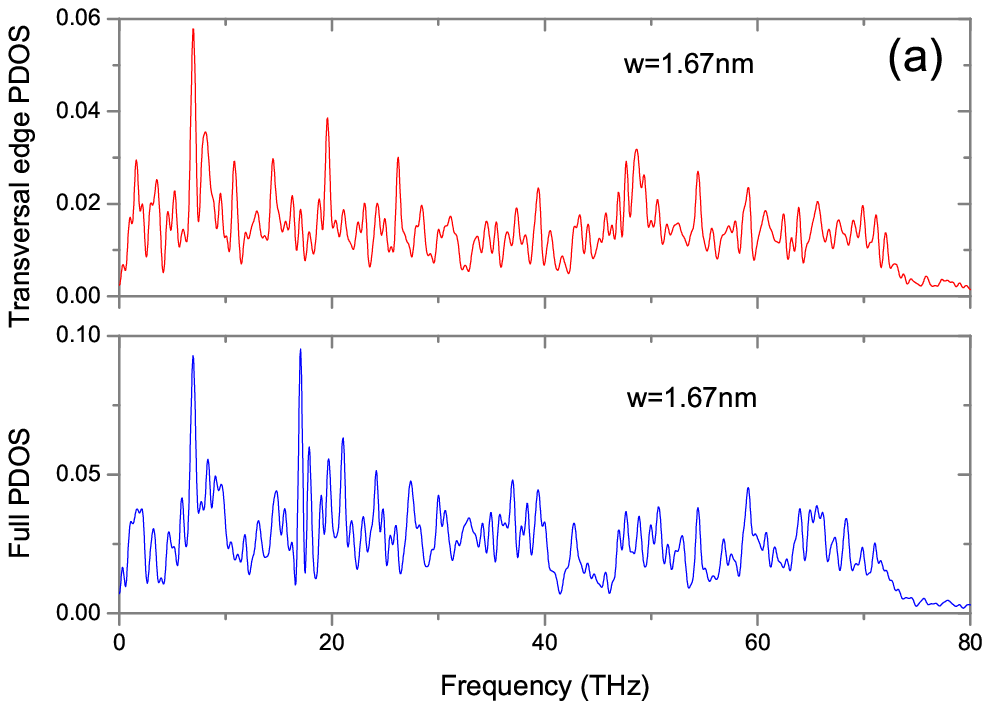}
\includegraphics[height=5cm]{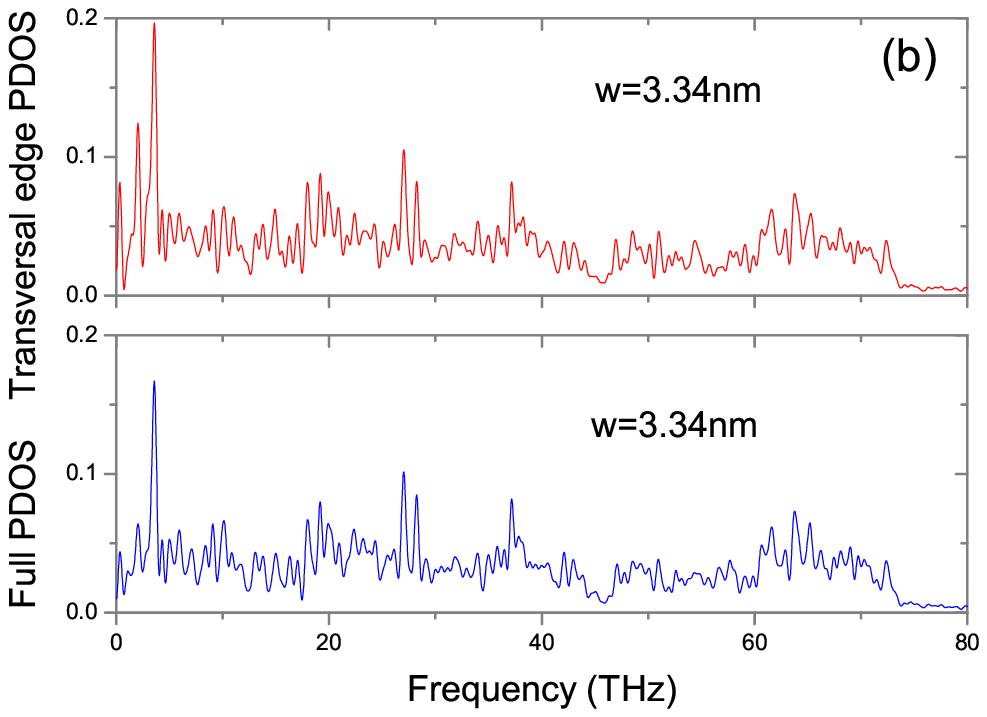}
\includegraphics[height=5cm]{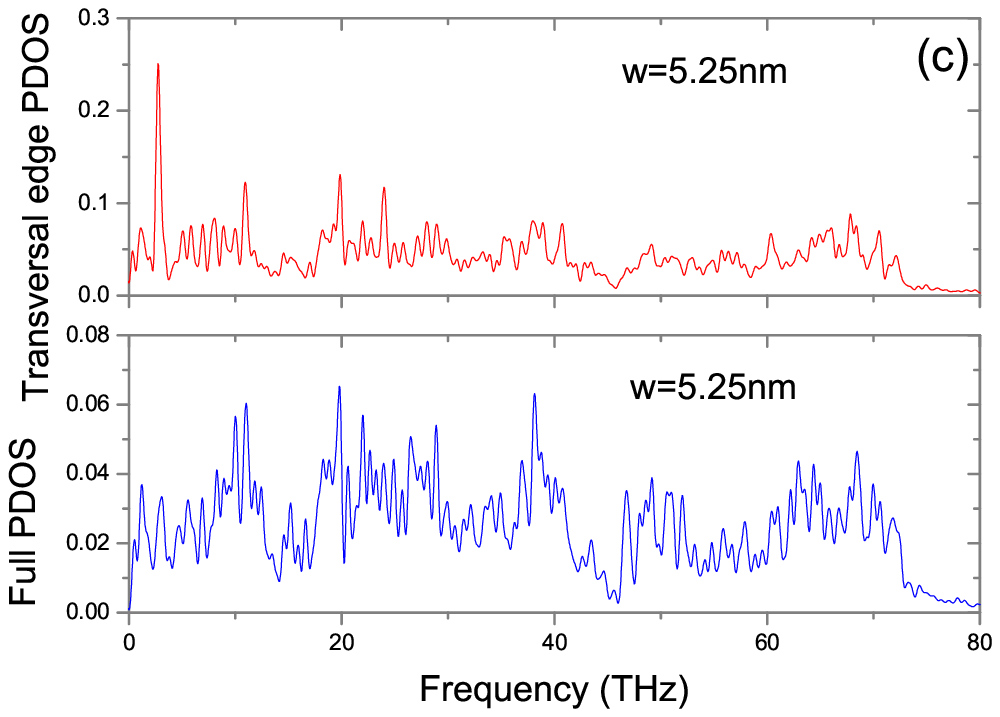}
\includegraphics[height=5cm]{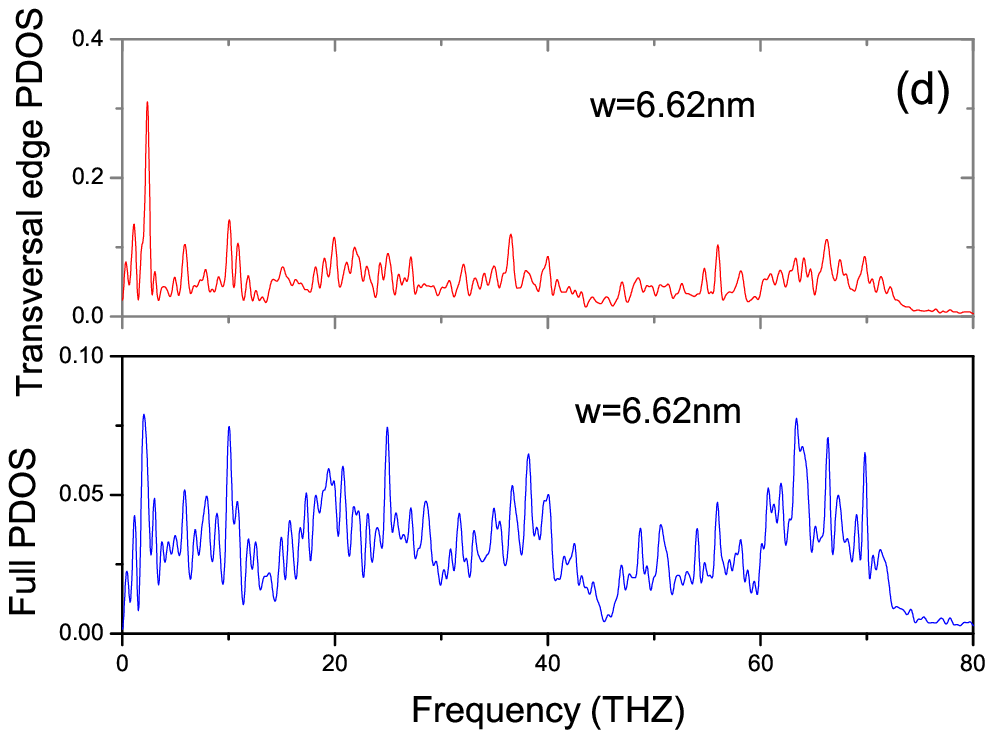}
\caption{(Color online) Full and transversal edge PDOS (arbitrary units) of single-layer armchair GNRs for different values of $w$ at $l=5nm$ and $\overline{T}=300K$. (a) $w=1.66nm$; (b) $w=3.34nm$; (c)$w=5.25nm$; (d)$w=6.62nm$.}\label{1}
\end{center}
\end{figure}

\indent In order to verify the analysis for Fig. 5, we also studied the width $w$ dependence of both full and transversal edge PDOS in the single-layer GNRs shown in Fig. 6.
For the transversal edge PDOS, the strong low-frequency peaks dominate the PDOS distribution for all values of the width (see the red line in Fig. 6).
 However, for full PDOS, the low-frequency peaks become gradually insignificant on increasing the width $w$.  For small width (e. g. $w=1.67nm$ and $3.34nm$), the low-frequency peaks in full PDOS are very prominent, the edge-localized phonon modes dominates the transport, so the thermal current reduces remarkably and NDTR can occur.  For large width (e. g. $w=5.25nm$ and $6.62nm$), the low-frequency peaks in full PDOS are not prominent, the effects of the edge-localized phonon modes reduce and NDTR disappears. Therefore, NDTR can not occur in very wide GNRs.

\begin{figure}[htbp]
\begin{center}\includegraphics[height=4.5cm]{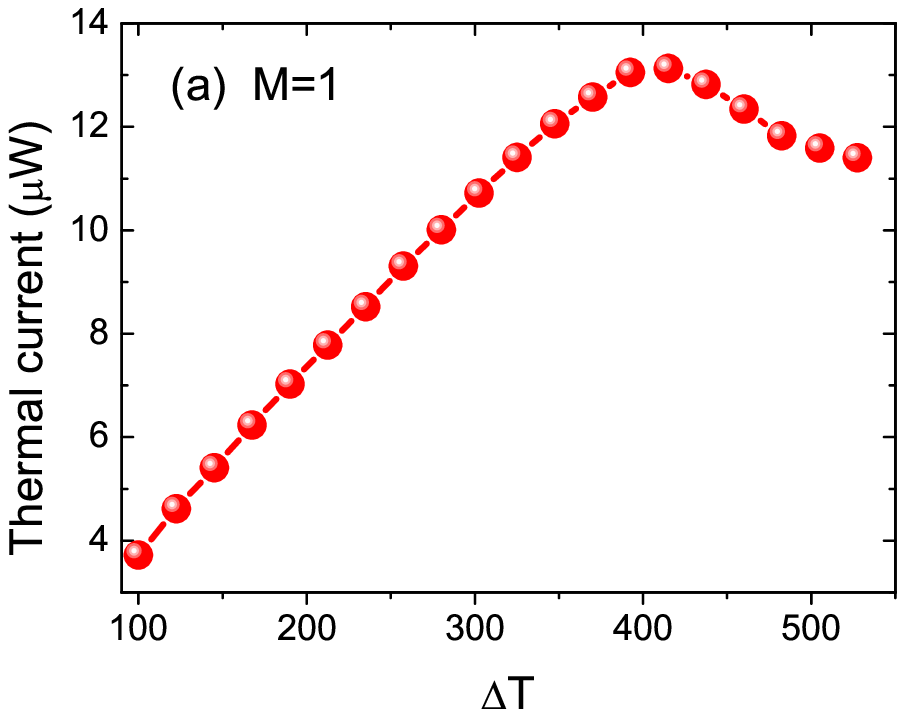}
\includegraphics[height=4.5cm]{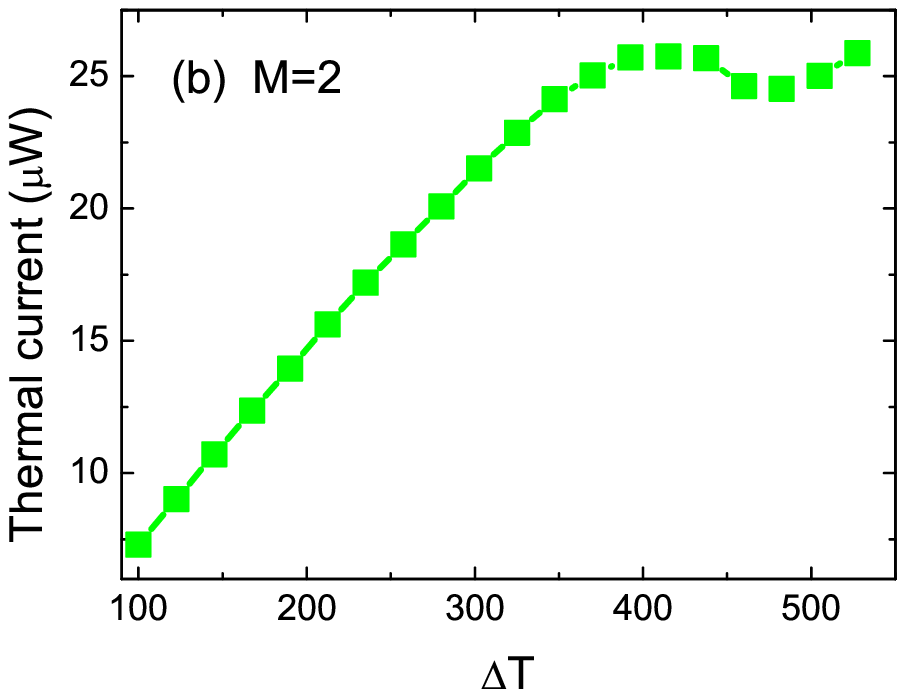}
\includegraphics[height=4.5cm]{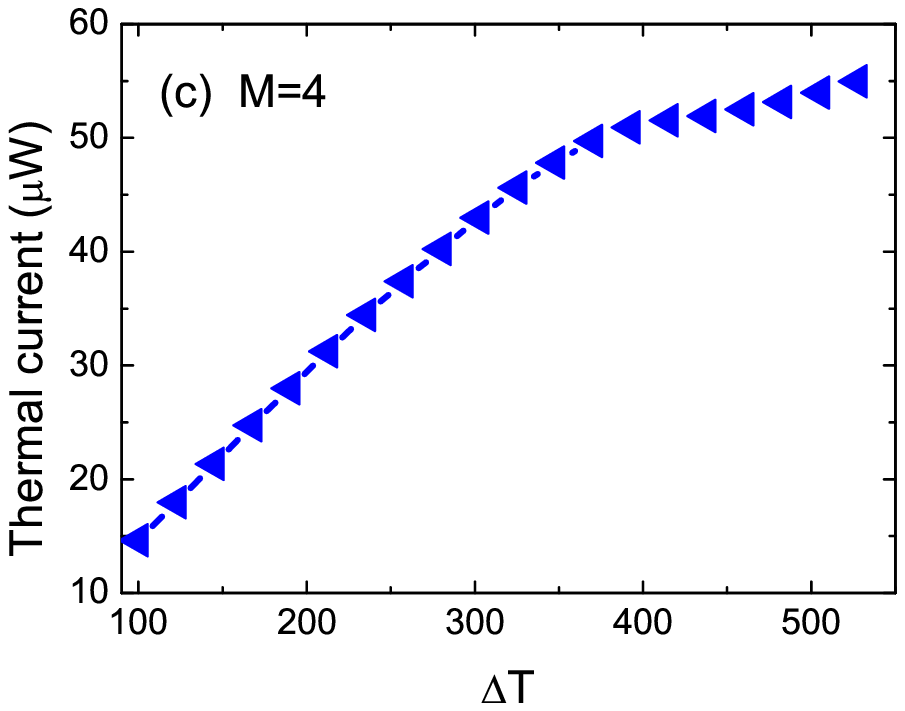}
\includegraphics[height=4.5cm]{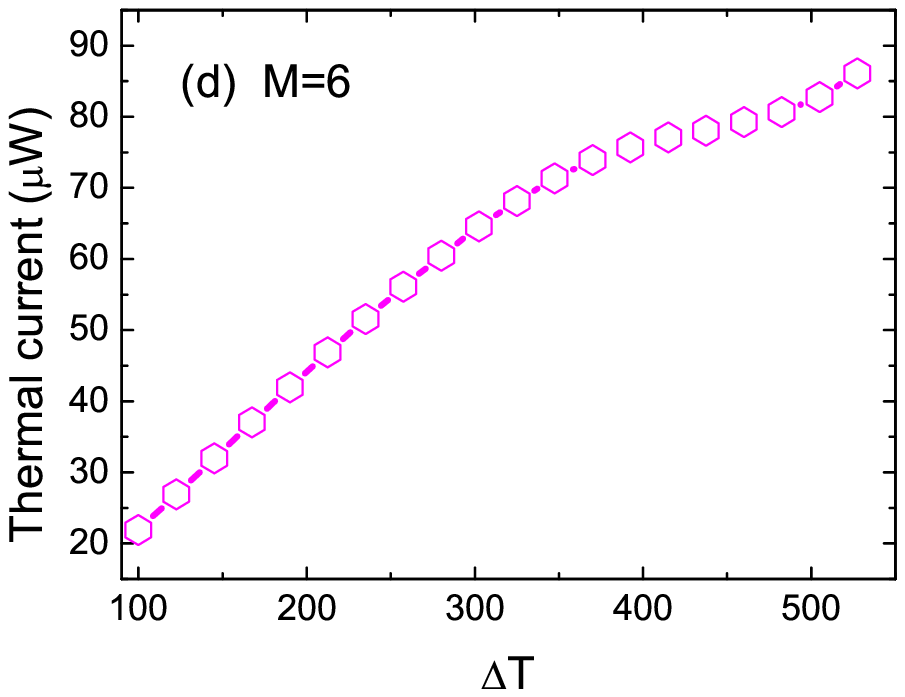}
\caption{(Color online) Thermal current $J$ as a function of temperature difference $\Delta T$ for different $M$ of armchair GNRs. (a)$M=1$; (b)$M=2$; (c)$M=4$; (d)$M=6$. $l=5nm$ and $w=1.67nm$. $T_{L}=600K$ and $T_{R}$ is varied.}\label{1}
\end{center}
\end{figure}

Figure 7 shows the layer dependence of NDTR for fixing $l=5nm$ and $w=1.67nm$.  It is found that NDTR regime becomes smaller as the number $M$ of the layers increases, and NDTR completely disappears for large $M$, e.g., $M=4, 6$.  This can be explained by the effect of the cross-plane coupling \cite{a16}.  For multi-layer GNRs, the cross-plane coupling will play an important role. In the presence of cross-plane coupling, the phonons will scatter with the atoms at the interface between the layers and then the effective thermal conductivity will decrease. This scattering effect from the cross-plane coupling decreases with the average temperature.
When $\Delta T$ increases from zero for fixing $T_{L}=600K$, the average temperature $\overline{T}$ will decrease and the scattering effect from cross-plane coupling reduces, therefore, the effective thermal conductivity increases and NDTR disappears.

\begin{figure}[htbp]
\begin{center}\includegraphics[height=7cm]{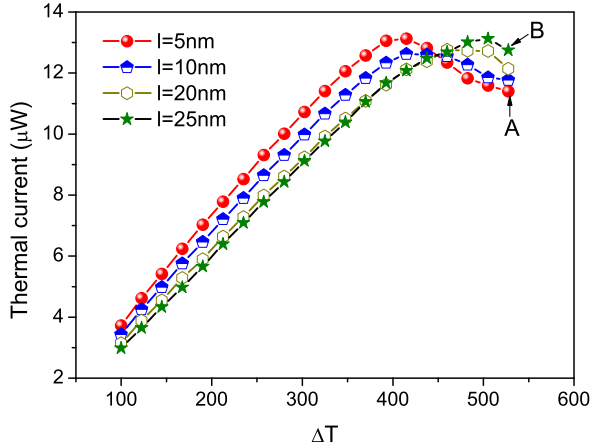}
\caption{(Color online) Thermal current $J$ as a function of the temperature difference $\Delta T$ for different lengths $l$ of single-layer armchair GNRs. $w=1.67$ and $M=1$. $T_{L}=600K$ and $T_{R}$ is varied. }\label{1}
\end{center}
\end{figure}

\begin{figure}[htbp]
\begin{center}\includegraphics[height=7cm]{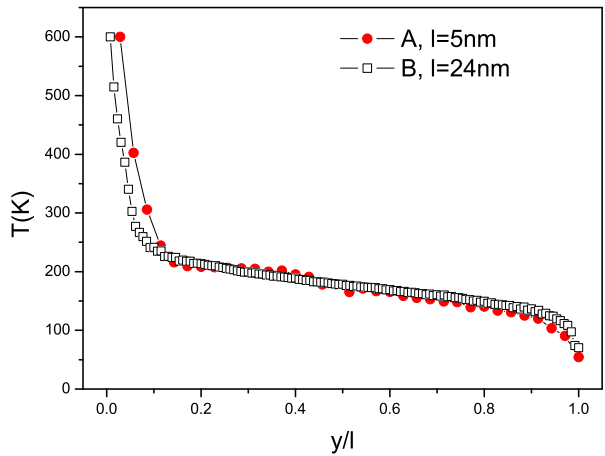}
\caption{Temperature profiles for different points $A$, $B$ described in Fig. (8). $y/l$ is the relative position along the transport.}\label{1}
\end{center}
\end{figure}

\indent Finally, we study the length dependence of NDTR in single-layer armchair GNRs.  From Fig. 8, we can find that the NDTR regime generally becomes smaller as the length increases.  For example, NDTR almost disappears for $l=25nm$. We thus suggest that NDTR will eventually disappear if the length exceeds some critical value.  Obviously,  when the length increases so that the length is much larger than the width ($l\gg w$),  GNRs becomes from two-dimensional system to one-dimensional system  and GNRs reduces to one-dimensional atom chain without on-site potential, where no NDTR can be found \cite{a14}.
On the other hand, we can check the temperature profiles of the point $A$ and $B$ shown in Fig. 8.  Though the shapes of the temperature profile for different length $l$ are similar (shown in Fig. 9), the temperature jump at high temperature boundary becomes small when the length $l$ increases.  So the thermal boundary resistance becomes not significant for very long GNRs and no NDTR occur.  Therefore, NDTR mainly occurs in small-size systems, which is in line with the current trend of device miniaturization in the technological world.
\section{Concluding remarks}
\indent In summary, we have investigated the thermal transport of GNRs in the nonlinear response regime from one dimension to three dimensions. When the width of the single-layer GNRs is very small, GNRs is a one-dimensional system, the onset of NDTR can not be observed.  On increasing the width $w$, GNRs becomes to a two-dimensional system, NDTR appears. However, for large width $w$, NDTR disappears.  When the number of the layers increases, GNRs becomes from two-dimensional system to three-dimensional system, NDTR regime becomes gradually smaller, and disappears for multiple layers (e.g., $M=4, 6$).  In addition, on increasing the length $l$ so that $l\gg w$, GNRs becomes from two dimensions  to one dimension, NDTR regime becomes smaller and eventually vanishes in the thermodynamic limit. The observation of NDTR in GNRs shows that NDTR can occur in a real system and GNRs may be good candidate materials for designing thermal devices.  Our results will give an important guidance for designing the graphene thermal devices where NDTR plays an important role. In addition, the study can also facilitate the understanding the onset of NDTR in low dimensional systems.

%\begin{figure}[htbp]
%\begin{center}\includegraphics[width=10cm,height=8cm]{fig3.eps}
%\caption{}\label{1}
%\end{center}
%\end{figure}

This work was supported in part by the National Natural Science Foundation
of China (Grant Nos.11004082 and 11175067), the Natural
Science Foundation of Guangdong Province, China (Grant
Nos.10451063201005249 and S201101000332) and the Fundamental Research Funds for the Central Universities,
JNU (Grant No. 21611437).

\end{document}